\documentclass[runningheads]{llncs}
\usepackage{fontawesome}  
\usepackage{cite}
\usepackage{amsmath,amssymb,amsfonts}

\usepackage[ruled, vlined, linesnumbered]{algorithm2e}

\SetCommentSty{mycommfont}
\usepackage{algpseudocode}
\usepackage{graphicx}
\usepackage{textcomp}
\usepackage{xcolor}
\usepackage{nicefrac}
\usepackage{hyperref}
\usepackage[font=scriptsize]{caption}

\def\BibTeX{{\rm B\kern-.05em{\sc i\kern-.025em b}\kern-.08em
    T\kern-.1667em\lower.7ex\hbox{E}\kern-.125emX}}

\newcommand{\orcid}[1]{
	\href{https://orcid.org/#1}{\includegraphics[scale=0.4]{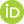}}
}

\newcommand{\mechanism}{\mathcal{M}}
\newcommand{\universeLog}{\mathcal{L}}
\newcommand{\universeActivity}{\mathcal{A}}
\newcommand{\universeEvent}{\mathcal{E}}

\begin{document}
\title{TraVaS: Differentially Private Trace Variant Selection for Process Mining\thanks{\scriptsize Funded under the Excellence Strategy of the Federal Government and the L{\"a}nder. We also thank the Alexander von Humboldt Stiftung for supporting our research.}}

\titlerunning{TraVaS: Differentially Private Trace Variant Selection for Process Mining}
% If the paper title is too long for the running head, you can set
% an abbreviated paper title here
%

\author{Majid Rafiei\orcid{0000-0001-7161-6927}\textsuperscript{\href{mailto:majid.rafiei@pads.rwth-aachen.de}{\faEnvelopeO}} \and
	Frederik Wangelik\orcid{0000-0001-6320-2302} \and
	Wil M.P. van der Aalst\orcid{0000-0002-0955-6940}}

\authorrunning{Majid Rafiei et al.}

\vspace{-0.3cm}
\institute{Chair of Process and Data Science, RWTH Aachen University, Aachen, Germany \\
% \email{\{majid.rafiei,wvdaalst\}@pads.rwth-aachen.de} 
% \email{gamal.elkoumy@ut.ee}
}
\maketitle              % typeset the header of the contribution
\vspace{-0.5cm}
\begin{abstract}
In the area of industrial process mining, privacy-preserving event data publication is becoming increasingly relevant. Consequently, the trade-off between high data utility and quantifiable privacy poses new challenges. State-of-the-art research mainly focuses on differentially private trace variant construction based on prefix expansion methods.
However, these algorithms face several practical limitations such as high computational complexity, introducing fake variants, removing frequent variants, and a bounded variant length.
In this paper, we introduce a new approach for direct differentially private trace variant release which uses anonymized \textit{partition selection} strategies to overcome the aforementioned restraints.
Experimental results on real-life event data show that our algorithm outperforms state-of-the-art methods in terms of both plain data utility and result utility preservation.

\end{abstract}

\section{Introduction}\label{c.intro}

In recent years, process mining and event data analysis have been successfully deployed in many industries. The main objectives are to learn process models from event logs for further behavioral inference (so-called \textit{process discovery}), to extend existing models using event logs (so-called \textit{model enhancement}), or to assess the alignment between a process model and an event log (so-called \textit{conformance checking})~\cite{book_wil}. 
% All of these application domains can support an organization to significantly streamline its workflows, optimize communication channels and improve the overall operating performance. 
However, often the underlying event data are bound to personal identifiers or other private information. A prominent example is the process management of hospitals where the cases are patients being treated by staff. 
% Table~\ref{tab:event_data} shows such an exemplary event log. 
Without means of privacy protection, any adversary is able to extract sensitive information about individuals and their properties. Thus, privacy regulations, such as GDPR~\cite{GDPR0}, typically restrict data storage and access which motivates the development of privacy preservation techniques.

\begin{table}[t]
\centering
\scriptsize
\caption{A simple event log from the healthcare context including trace variants and their frequencies.}
\label{tab:trace_variant}
    \begin{tabular}{l|c}
    \hline
    Trace Variant   & Frequency \\ 
    \hline
    $\langle register, visit, blood\text{-}test, release \rangle$                    & 10 \\ 
    $\langle register, blood\text{-}test, visit, release \rangle$                    & 8 \\
    $\langle register, visit, release \rangle$                                       & 20 \\ 
    $\langle register, visit, blood\text{-}test, blood\text{-}test, release \rangle$ & 5  \\ \hline
    \end{tabular}
    \vspace{-0.5cm}
\end{table}

The majority of state-of-the-art privacy preservation techniques are built on Differential Privacy (DP), which offers a noise-based privacy definition.
This is due to its important features, such as providing mathematical privacy guarantees and security against \textit{predicate-singling-out} attacks \cite{PSO}. 
% In this respect, Differential Privacy (DP), which provides a noise-based definition for privacy, is becoming the basis for most of state-of-the-art techniques. That is because of its important features such as providing mathematical privacy guarantees and security against \textit{predicate-singling-out} attacks \cite{PSO}. 
The goal of techniques based on DP is to hide the participation of an individual in the released output by injecting noise.
The amount of noise is mainly determined by the privacy parameters, $\epsilon$ and $\delta$, and the sensitivity of the underlying data. 
State-of-the-art research targeting ($\epsilon, \delta$)-DP methods in process mining focuses on releasing raw privatized activity sequences performed for cases, i.e., \textit{trace variants}.
Table~\ref{tab:trace_variant} shows a sample of such event data in the healthcare context, where each trace variant belongs to a case, i.e., a patient, and one case cannot have more than one trace variant. 
This format describes the \textit{control-flow} of event logs that is basis for the main process mining activities.
The trace variant of a case is considered sensitive information because it contains the complete sequence of activities performed for the case that can be exploited to conclude private information, e.g., patient diseases in the healthcare context.
% Consider, for instance, the healthcare context where the activities are treatment-related, a sequence of such activities can be exploited to determine the private health information of patients, e.g., their disease. 
% For instance, in the simple event log shown in Table~\ref{tab:trace_variant}, one may know that two blood tests can only done for specific type of diseases. 

To achieve differential privacy for trace variants, the state-of-the-art approach \cite{priv_pinq} inserts noise drawn from a \textit{Laplacian distribution} into the variant distribution obtained from an event log.
This approach has several drawbacks including: (1) \textit{introducing fake variants}, (2) \textit{removing frequent true variants}, and (3) \textit{limited length for generated trace variants}. 
A recent work called \emph{SaCoFa} \cite{priv_sacofa}, attempts to mitigate drawbacks (1) and (2) by gaining knowledge regarding the underlying process semantics from original event data.
However, the privacy quantification of all extra queries to gain knowledge regarding the underlying semantics is not discussed. 
Moreover, the third drawback still remains since this work, similar to \cite{priv_pinq}, employs a \textit{prefix-based} approach. The prefix-based approaches need to generate all possible unique variants based on a set of activities to provide differential privacy for the original distribution of variants. 
Since the set of possible trace variants that can be generated given a unique set of activities is infinite, the prefix-based techniques need to bound the length of generated sequences. Also, to limit the search space these approaches typically include a pruning parameter to exclude less frequent prefixes.

% \begin{figure}[t]
% \centering
% \includegraphics[width=0.49\textwidth]{Intro.jpg}
% \caption{Schematic view of different ($\epsilon, \delta$)-DP event log transformations. The upper part represents the main cycle of iterative tree-based (TB) methods while the principle of partition selection (PS) is shown in the lower half. Both approaches access a private event log and release anonymized versions. $\sigma = \langle a_1,a_2,\dots,a_n \rangle ^n$ denotes a trace variant occurring $n$ times within the log. $f_i \in \mathbb{N}_{>0}$ denotes the actual frequency of $\sigma_i$, and $k$ denotes a frequency threshold. $x_i \in \mathbb{Z}$ denotes the noise that is symmetric at 0.}\label{fig1}
% \end{figure}

We introduce an ($\epsilon, \delta$)-DP approach for releasing the distribution of trace variants that focuses on the aforementioned drawbacks. In contrast to the prefix-based approaches, the underlying algorithm is based on ($\epsilon, \delta$)-DP for \textit{partition selection} that allows for a direct publication of arbitrarily long sequences \cite{priv_part}.
Employing differentially private partition selection techniques, the actual frequencies of all trace variants can directly be queried without guessing (generating) trace variants. Internally, random noise drawn from a specific geometric distribution is injected into the corresponding frequencies, and all variants whose privatized frequencies fall beyond a threshold are removed. Hence, no fake trace variants are introduced, and only some infrequent variants may disappear from the output. Moreover, no tedious fine-tuning has to be conducted and no computationally expensive search needs to be included.
In Section~\ref{c.exp}, we introduce different metrics to evaluate the \textit{data} and \textit{result} utility preservation of our approach. We also run our experiments for the state-of-the-art prefix-based methods and show superior data and result utilities compared to these methods.

The remainder of this paper is structured as follows. In Section~\ref{c.relwork}, we provide a summary of related work. Preliminaries and notations are provided in Section~\ref{c.prelim}. Section~\ref{c.partsel} introduces the theoretical background of differentially private \textit{partition selection}, and describes our \emph{TraVaS} algorithm. In Section~\ref{c.exp}, the experimental results based on real-life event logs are shown. Section~\ref{c.conc} concludes the paper.

\section{\label{c.relwork}Related Work}

The research area of privacy and confidentiality in process mining is recently growing in importance. 
Several techniques have been proposed to address the privacy and confidentiality issues.
In this paper, our focus is on the so-called \textit{noise-based} techniques that are based on the notion of \textit{differential privacy}.
In \cite{priv_pinq}, the authors apply an ($\epsilon, \delta$)-DP mechanism to event logs to privatize \textit{directly-follows relations} and trace variants. 
The underlying principle uses a combination of an ($\epsilon, \delta$)-DP noise generator and an iterative query engine that allows an anonymized publication of trace variants with an upper bound for their length.
\emph{SaCoFa} \cite{priv_sacofa} is the most recent extension of the aforementioned ($\epsilon, \delta$)-DP mechanism that attempts to optimize the query structures with the help of underlying semantics. 
Another extension of \cite{priv_pinq} is the \textit{PRIPEL} approach, where more event attributes can be secured using the so-called \textit{sequence enrichment} \cite{priv_pripel}. 

% Moreover, in \cite{priv_cont}, researchers demonstrate how \textit{group-based} models can deter correspondence attacks against continuous event data publishing.

Whereas most of the aforementioned ideas target raw event logs, in \cite{priv_graph}, the focus is on \textit{directly-follows graphs}. During the edge generation, connections are randomized using ($\epsilon, \delta$)-DP mechanisms to balance utility preservation and privacy risks.
As the main benchmark model for our work, we choose the technique by Mannhardt et al. \cite{priv_pinq} since it focuses on trace variants and is the basis of most of the other techniques. 
Moreover, its privacy guarantees are directly proven by ($\epsilon, \delta$)-DP mechanisms, i.e., no extra privacy analysis is required. Nevertheless, we also compare our results with SaCoFa as the most recent extension of the benchmark to demonstrate the superior performance of our approach.

% There are also privacy preservation techniques in process mining that adapted the notion of $k$-anonymity and its extensions such as \cite{priv_pretsa} and \cite{rafiei_group}.  In \cite{priv_pretsa}, the authors offer \textit{k-anonymity} and \textit{t-closeness} sanitizers to avoid disclosing the identities of \textit{resources}, i.e., activity performers. 
% In \cite{priv_ind}, a uniformization-based approach is introduced to preserve privacy of cases during process mining analyses.
% In \cite{rafiei_group}, the TLKC-privacy model is proposed as another group-based privacy preservation technique which utilizes \textit{k-anonymity} components to protect cases in event logs. 

% Although the suppression results were promising, \textit{k-anonymity} mechanisms do not entirely mitigate PSO attacks.

% In addition, the resulting anonymized output can be directly understood by most mining algorithms.
% and the corresponding implementation is flexible to use.

\section{\label{c.prelim}Preliminaries}

In this section, we introduce the necessary mathematical concepts and definitions utilized throughout the remainder of the paper.
% The data formats used in the following definitions are strongly linked to basic set operations which will be briefly highlighted.
Let $A$ be a set. $B(A)$ is the set of all multisets over $A$. 
A multiset $A$ can be represented as a set of tuples $\{ (a,A(a)) | a \in A \}$ where $A(a)$ is the frequency of $a \in A$.
Given $A$ and $B$ as two multisets, $A \uplus B$ is the sum over multisets, e.g., $[a^2,b^3] \uplus [b^2,c^2] = [a^2,b^5,c^2]$.
We define a finite sequence over $A$ of length $n$ as $\sigma {=} \langle a_1, a_2,\dots, a_n\rangle$ where $\sigma(i) {=} a_i {\in} A$ for all $i {\in} \{1,2,\dots,n\}$. The set of all finite sequences over $A$ is denoted with $A^*$. 
% and the set of all elements of $\sigma$ is written as $\{a {\in} \sigma\}$.

\vspace{-0.2cm}
\subsection{\label{sub.event}Event Data}
The data used by \textit{process mining} techniques are typically collections of unique events that are recorded per activity execution and characterized by their attributes. 
We denote $\universeEvent$ as the universe of events. Then, a \textit{trace} $\sigma$, which is a single process execution, is represented as a sequence of events $\sigma {=} \langle e_1,e_2,...,e_n \rangle \in \universeEvent^*$ belonging to the same case and having a fixed ordering based on timestamps.
Note that events are unique and cannot appear in more than one trace. Moreover, each case (individual) contributes to only one trace.
An event log $L$ can be represented as a set of traces $L {\subseteq} \universeEvent^*$. Our work focuses on the control-flow aspect of an event log that only considers the activity attribute of events in traces.
We define a simple event log based on activity sequences, so-called \textit{trace variants}.

\begin{definition}[Trace Variant]\label{def:trace_var}
Let $\universeActivity$ be the universe of activities. A trace variant $\sigma = \langle a_1,a_2,...,a_n \rangle \in \universeActivity^*$ is a sequence of activities performed for a case.
\end{definition}

\begin{definition}[Simple Event Log]\label{def:simple_el}
A simple event log $L$ is defined as a multiset of trace variants $L \in B(\universeActivity^*)$. $\universeLog$ denotes the universe of simple event logs.
\end{definition}

\subsection{\label{sub.priv}Differential Privacy}
% The most prominent privacy preservation techniques to date are \textit{group-based} techniques based on \textit{k-anonymity} and its extensions as well as ($\epsilon, \delta$)-DP mechanisms \cite{priv_dwork2}. Although both approaches have demonstrated superior performance compared to former competitors, ($\epsilon, \delta$)-DP gains more and more traction due to its ability to accurately measure privacy loss and to guarantee security against deanonymization attacks \cite{priv_dwork2}.

In the following, we introduce the necessary concepts of $(\epsilon,\delta)$-DP for our research.
The main idea of DP is to inject noise into the original data in such a way that an observer who sees the randomized output cannot tell if the information of a specific individual is included in the data \cite{priv_dwork2}.
Considering simple event logs, i.e., the distribution of trace variants, as our sensitive event data,
differential privacy can formally be defined as Definition~\ref{def:dp}.

% \begin{definition}[Differential Privacy]\label{def:dp}
% Let $D_1$ and $D_2$ be two neighboring tabular databases that differ only in a single row entry. Further let $\mathcal{M}$ be a randomized mechanism which takes a database as input. Then $\mathcal{M}$ is said to provide $(\epsilon, \delta)$-differential privacy if for all $ S \subseteq rng(\mathcal{M})$, $Pr[\mathcal{M}(D_1) \in S] \leq e^\epsilon \times Pr[\mathcal{M}(D_2) \in S] + \delta$. 
% \end{definition}

% -----Detailed definition----
\begin{definition}[($\epsilon$,$\delta$)-DP for Event Logs]\label{def:dp}
Let $L_1$ and $L_2$ be two neighbouring event logs that differ only in a single entry, e.g., $L_2 {=} L_1 {\uplus} [\sigma]$ for any $\sigma {\in} \universeActivity^*$.
Also, let $\epsilon {\in} \mathbb{R}_{>0}$ and $\delta {\in} \mathbb{R}_{>0}$ be two privacy parameters. 
A randomized mechanism $\mechanism_{\epsilon,\delta}{:} \universeLog {\to} \universeLog$ provides ($\epsilon,\delta$)-DP if for all $S {\subseteq} \universeActivity^* {\times} \mathbb{N}$: 
$Pr[\mechanism_{\epsilon,\delta}(L_1) \in S] \leq e^\epsilon {\times} Pr[\mechanism_{\epsilon,\delta}(L_2) \in S] {+} \delta$. 
Given $L \in \universeLog$, $\mechanism_{\epsilon,\delta}(L) \subseteq \{ (\sigma,L'(\sigma)) \mid \sigma \in \universeActivity^* \wedge L'(\sigma) = L(\sigma) + x_{\sigma} \}$, with $x_{\sigma}$ being realizations of i.i.d. random variables drawn from a probability distribution.
\end{definition}

In Definition~\ref{def:dp}, $\epsilon$ as the first privacy parameter specifies the probability ratio, and $\delta$ as the second privacy parameter allows for a linear violation. In the strict case of $\delta = 0$, $\mechanism$ offers $\epsilon$-DP.
The randomness of respective mechanisms is typically ensured by the noise drawn from a probability distribution that perturbs original variant-frequency tuples and results in non-deterministic outputs. 
The smaller the privacy parameters are set, the more noise is injected into the mechanism outputs, entailing a decreasing likelihood of tracing back the instance existence based on outputs.

A commonly used $(\epsilon, 0)$-DP mechanism for real-valued statistical queries is the \textit{Laplace} mechanism. This mechanism injects noise based on a Laplacian distribution with scale $\nicefrac{\Delta f}{\epsilon}$.  
$\Delta f$ is called the sensitivity of a statistical query $f$.
Intuitively, $\Delta f$ indicates the amount of uncertainty we must introduce into the output in order to hide the contribution of single instances at $(\epsilon, 0)$-level. 
In our context, $f$ is the frequency of a trace variant. Since one individual, i.e., a case, contributes to only one trace, $\Delta f {=} 1$. In case an individual can appear in more than one trace, the sensitivity needs to be accordingly increased assuming the same value for the privacy parameter $\epsilon$.
State-of-the-art event data anonymization frameworks such as our benchmark often use the \textit{Laplace mechanism}.

% One important property of $(\epsilon, \delta)$-DP is its behavior under compositional deployment. In the following, we provide the \textit{mechanism composition theorem} taken from \cite{priv_dwork2}. 

% As Theorem 1 states, $(\epsilon, \delta)$-DP mechanisms can be combined into more complex algorithms at the cost of a directly measurable cumulative privacy loss. Nevertheless, the result still promises $(\epsilon, \delta)$-DP independent of the exact form of composition or query structure.

% We use the decomposition and post-proccessing immunity properties of $(\epsilon, \delta)$-DP mechanisms in our \textit{TraVas optimizer} algorithm (Algorithm~\ref{algo.mult}). 

\section{\label{c.partsel}Partition Selection Algorithm}

We first highlight the problem of \textit{partition selection} and link it to event data release. Then, the algorithmic details are presented with a brief analysis.

% \vspace{-0.2cm}
\subsection{Partition Selection}

Many data analysis tasks can be expressed as per-partition aggregation operations after grouping the data into an unbounded set of partitions. When identifying the variants of a simple log $L$ as categories, the transformation from $L$ to pairs ($\sigma, L(\sigma)$) becomes a specific instance of these aggregation tasks.
To render such queries differentially private, two distinct steps need to be executed. First, all aggregation results are perturbed by noise addition of suitable mechanisms. Next, the set of unique partitions must be modified to prevent leakage of information on the true data categories (\textit{differentially private partition selection}) \cite{priv_dwork2,priv_part}. In case of publicly known partitions or bounded partitions from a familiar finite domain, the second step can be reduced to a direct unchanged release or a simple guessing-task, respectively. However, for the most general form of unknown and infinite category domains, guessing is not efficient anymore and an $(\epsilon, \delta)$-DP \textit{partition selection} strategy can be used instead.

Recently, in \cite{priv_part}, the authors proposed an $(\epsilon, \delta)$-DP \textit{partition selection} approach, where they provided a proof of an optimal partition selection rule which maximizes the number of released category-aggregation pairs while preserving $(\epsilon, \delta)$-DP.
In particular, the authors showed how the aforementioned anonymization steps can be combined into an explicit $(\epsilon, \delta)$-DP mechanism based on a k-Truncated Symmetric Geometric Distribution (k-TSGD), see Definition~\ref{def:k_tsgd}.
We exploit the analogy between \textit{partition selection} and simple event log publication and transfer this mechanism to the event data context. Definition~\ref{def:dp_partition} shows the respective definition based on a k-TSGD.\footnote{\scriptsize A respective proof can be found in Sec. 3 of \cite{priv_part}.}

% Recently, the first corresponding approach was developed as an answer to the private \textit{set union} problem \cite{priv_part}. Following research then adapted the main concepts and provided a proof of an optimal \textit{partition selection} rule which maximizes the number of released category-aggregation pairs while still preserving $(\epsilon, \delta)$-DP \cite{priv_part}. 

\begin{definition}[k-TSGD]\label{def:k_tsgd}
    Given probability $p {\in} (0, 1)$, $m {=} \nicefrac{p}{(1+(1 - p)-2(1-p)^{k+1})}$, and $k \geq 1$, the k-TSGD of $(p, k)$ over $\mathbb{Z}$ formally reads as:
    \begin{equation}
    \footnotesize
    \text{k-TSGD}[ X = x\ |\ p, k] = 
        \begin{cases}
          m \cdot (1 - p)^{|x|}  & \text{if $x \in [-k, k]$}\\
          0 & \text{otherwise}
        \end{cases}       
    \end{equation}
\end{definition}

\begin{definition}[($\epsilon$,$\delta$)-DP for Event Logs Based on k-TSGD]\label{def:dp_partition}
    Let $\epsilon {\in} \mathbb{R}_{>0}$ and $\delta {\in} \mathbb{R}_{>0}$ be the privacy parameters, and $\mechanism_{\epsilon,\delta}^{k-TSGD}: \universeLog \rightarrow{\universeLog}$ be a randomized mechanism based on a k-TSGD.
    Given $L \in \universeLog$ as an input of the randomized mechanism, an event log $L'{=} \{(\sigma, L'(\sigma)) \mid \sigma {\in} L \wedge L'(\sigma) {>} k\} \in rng(\mechanism_{\epsilon,\delta}^{k-TSGD})$ is an $(\epsilon, \delta)$-DP representation of $L$ if 
    $L'(\sigma) {=} L(\sigma) {+} x_{\sigma}$ is the noisified frequency with $x_{\sigma}$ being realization of i.i.d random variables drawn from a k-TSGD with parameters $(p, k)$, where
    $p = 1 {-} e^{-\epsilon}$ and 
    $k = \lceil \nicefrac{1}{\epsilon} {\times} ln ( \nicefrac{(e^\epsilon + 2\delta - 1)}{\delta(e^\epsilon + 1)}) \rceil$.
\end{definition}

Definition~\ref{def:dp_partition} shows the direct $(\epsilon, \delta)$-DP release of trace variants by first perturbing all variant frequencies and then truncating infrequent behavior. Additionally, optimality is guaranteed w.r.t. the number of variants being published due to the \textit{k-TSGD} structure \cite{priv_part}. Note that the underlying \textit{k-TSGD} mechanism assumes each case only contributes to one variant. In case this requirement needs to be violated, sensitivity considerations force a decrease in $(\epsilon, \delta)$. 
% or a split into sublogs to achieve comparable privacy levels.

The development of differentially private \textit{partition selection} enables significant performance improvements for private trace variant releases. As there are infinite activity sequences defining a variant, former approaches had to either guess or query all of these potentially non-existing sequences in a cumbersome fashion due to the ex-ante category anonymity in $(\epsilon, \delta)$-DP. 
% Examples are \cite{priv_pinq} and \cite{priv_sacofa} where the authors employed iteratively built prefix trees of activity sequences to reduce the brute force workload and retrieve corresponding variant frequencies \cite{priv_pinq}. 
On the contrary, \textit{partition selection} only needs one noisified aggregation operation followed by a specific truncation. Hence, the output contains only existing variants that are independent of external parameters or query patterns. 

% \vspace{-0.2cm}
\subsection{Algorithm Design}
Algorithm~\ref{algo.single} presents the core idea of \textit{TraVaS} which is based on Definition~\ref{def:dp_partition}. We also propose a utility-aware extension of \textit{TraVaS}, so-called \textit{uTraVaS}, that utilizes the privacy budgets, i.e., $\epsilon$ and $\delta$, by several queries w.r.t. data utility. In this paper, we focus on \textit{TraVaS}, the details of \textit{uTraVaS} are provided on GitHub.\footnote{\scriptsize \url{https://github.com/wangelik/TraVaS/tree/main/supplementary}}   

% The goal of our privatization algorithm is to convert Definition~\ref{def:ktsgd} into an efficient procedure that gets a confidential log $L$ and optimizes the privatized result w.r.t. data utility.
% Since such an algorithm must have access to the original data, unauthorized entities should only interact with the anonymized output and not observe internal data flows.

% Our implementation consists of two main parts: an $(\epsilon, \delta)$-DP privatization core and a multi-query optimizer shell. The core is depicted as Algorithm~\ref{algo.single} (\textit{SQVR}) and allows to anonymize variant-frequency pairs by injecting \textit{k-TSGD} noise within one run over the according simple log. 

\begin{algorithm}[b]
\scriptsize
\DontPrintSemicolon
\KwIn{Event log $L$, DP-Parameters ($\epsilon, \delta$)}
\KwOut{($\epsilon, \delta$)-DP log $L'$}
\SetKwBlock{Begin}{function}{end function}
\Begin(\text{travas} ${(} L, \epsilon, \delta{)}$)
{
  $p = 1-e^{-\epsilon}$ \tcp*{compute probability}
  $k = \lceil 1/\epsilon $ $ \times $ ln $ ((e^\epsilon+2\delta-1) / (\delta(e^\epsilon+1))) \rceil$   \tcp*{compute threshold}
  \ForAll{$(\sigma, L(\sigma)) \in L$}
  {
    $x_{\sigma} = $ rTSGD $ \left( p, k\right)$   \tcp*{generate i.i.d k-TSGD noise}
    \uIf{$L(\sigma) + x_{\sigma} > k$}
    {
        add $(\sigma,L(\sigma)+x_\sigma)$ to $L'$
    }
    
  }
  \Return{$L'$}
}
\caption{Differentially Private Trace Variant Selection (TraVaS)}\label{algo.single}
\end{algorithm}

Algorithm~\ref{algo.single} (TraVaS) allows to anonymize variant-frequency pairs by injecting \textit{k-TSGD} noise within one run over the according simple log. 
After a simple log $L$ and privacy parameters $(\epsilon > 0, \delta > 0)$ are provided, the \textit{travas} function first transforms $(\epsilon, \delta)$ into \textit{k-TSGD} parameters $(p, k)$. 
Then, each variant frequency $L(\sigma)$ becomes noisified using i.i.d \textit{k-TSGD} noise $x_{\sigma}$ (see Definition~\ref{def:dp_partition}). Eventually, the function removes all modified infrequent variants where the perturbed frequencies yield numbers below or equal to $k$.
Due to the partition selection mechanism, the actual frequencies of all trace variants can directly be queried without guessing trace variants. Thus, \textit{TraVaS} is considerably more efficient and easier to implement than current state-of-the-art prefix-based methods.

\section{\label{c.exp} Experiments}

We compare the performance of \textit{TraVaS} against the state-of-the-art benchmark \cite{priv_pinq} and its extension (\emph{SaCoFa} \cite{priv_sacofa}) on real-life event logs. 
% These methods achieve $(\epsilon, \delta)$-DP by guessing valid variants with the help of iteratively expanding activity sequences. 
Due to algorithmic differences between our approach and the prefix-based approaches, it is particularly important to ensure a fair comparison.
Hence, we employ divergently structured event logs and study a broad spectrum of privacy budgets $(\epsilon, \delta)$. Moreover, the sequence cutoff for the benchmark and \emph{SaCoFa} is set to the length that covers 80\% of variants in each log, and the remaining pruning parameter is adjusted such that on average anonymized logs contain a comparable number of variants with the original log.
Note that \textit{TraVaS} guarantees the optimal number of output variants due to its underlying differentially private partition selection mechanism \cite{priv_part}, and it does not need to limit the length of the released variants. 
Thus, the aforementioned settings consider the limitations of the prefix-based approaches to have a fair comparison.

% Such sequences are organized as tree-like structures of prefixes, and their frequencies are queried from the original event log. Additionally, a maximum sequence length is defined and infrequent occurrences are pruned to limit the exponential growth.

% Due to the algorithmic similarities between our benchmark and \emph{SaCoFa}, the privacy quantification limitations of SaCofa explained in Sec.~\ref{c.intro}, and the space limitation, we mainly focus on our benchmark. Nevertheless, the detailed results of \emph{SaCoFa} are also available in our GitHub repository and are briefly discussed.

\begin{table}[b]
\vspace{-0.8cm}
\scriptsize
\centering
\caption{General statistics of the event logs used in our experiments.}
\label{tab:exp_data}
\begin{tabular}{c|c|c|c|c|c}
\hline
Event Log & \#Events & \#Cases & \#Activities & \#Variants & Trace Uniqueness\\
\hline
Sepsis & 15214 & 1050 & 16 & 846 & 80\%\\
BPIC13 & 65533 & 7554 & 4 & 1511 & 20\%\\
\hline
\end{tabular}
\end{table}

We select two event logs of varying size and trace uniqueness. As we discussed in Section~\ref{c.partsel}, and it is considered in other research such as \cite{priv_pinq}, \cite{priv_sacofa}, and \cite{rafiei_group}, infrequent variants are challenging to privatize. Thus, trace uniqueness is an important analysis criterion.
The Sepsis log describes hospital processes for Sepsis patients and contains many rare traces \cite{data_sepsis}. 
In contrast, BPIC13 has significantly more cases at a four times smaller trace uniqueness \cite{data_bpic}. The events in BPIC13 belong to an incident and problem management system called VINST. Both logs are realistic examples of confidential human-centered information where the case identifiers refer to individuals.
Detailed log statistics are shown in Table~\ref{tab:exp_data}.

% \vspace{-0.2cm}
\subsection{\label{s.metrics} Evaluation Metrics}

To assess the performance of an $(\epsilon, \delta)$-DP mechanism, suitable evaluation metrics are needed to determine how valuable the anonymized outputs are w.r.t. the original data. 
In this respect, we first consider a \textit{data utility} perspective where the similarity between two logs is measured independent of future applications. For our experiments, two respective metrics are considered. 
From \cite{priv_emd}, we adopt \textit{relative log similarity} that is based on the \textit{earth mover's distance} between two trace variant distributions, where the normalized \textit{Levenshtein} string edit distance is used as a similarity function between trace variants.
The \textit{relative log similarity} metric quantifies the degree to which the variant distribution of an anonymized log matches the original variant distribution on a scale from 0 to 1. 

In addition, we introduce an \textit{absolute log difference} metric to account for situations where distribution-based metrics provide only different expressiveness. Exemplary cases are event logs possessing similar variant distributions, but significantly different sizes. For such scenarios, the \emph{relative log similarity} yields high similarity scores, whereas \emph{absolute log difference} can detect these size disparities.
To derive an absolute log difference value, we first transform both input logs into a \textit{bipartite graph} of variant vertices. Then a \textit{cost network flow} problem \cite{cnf} is solved by setting demands and supplies to the absolute variant frequencies and utilizing a \textit{Levenshtein} distance between variants as an edge cost. 
Hence, the resulting optimization value of an $(\epsilon, \delta)$-DP log resembles the number of \textit{Levenshtein} operations to transform all respective variants into variants of the original log. In contrast to our \textit{relative log similarity} metric, this approach can also penalize a potential matching impossibility. More information on the exact algorithms is provided on GitHub.\footnote{\scriptsize \url{https://github.com/wangelik/TraVaS/tree/main/supplementary}}

Besides comparing event logs based on \emph{data utility} measures, we additionally quantify the algorithm performance with \textit{process discovery} oriented \textit{result utilities}. We use the \textit{inductive miner infrequent} \cite{mine_infreq} with default noise threshold of 20\% to discover process models from the privatized event logs for all $(\epsilon, \delta)$ settings under investigation. Then, we compare the models with the original event log to obtain token-based replay \textit{fitness} and \textit{precision} scores \cite{book_wil}.
Due to the probabilistic nature of $(\epsilon, \delta)$-DP, we average all metrics over 10 anonymized logs for each setting, i.e., 10 separate algorithm runs per setting.

\begin{figure}[t]
\centering
\includegraphics[width=0.95\textwidth]{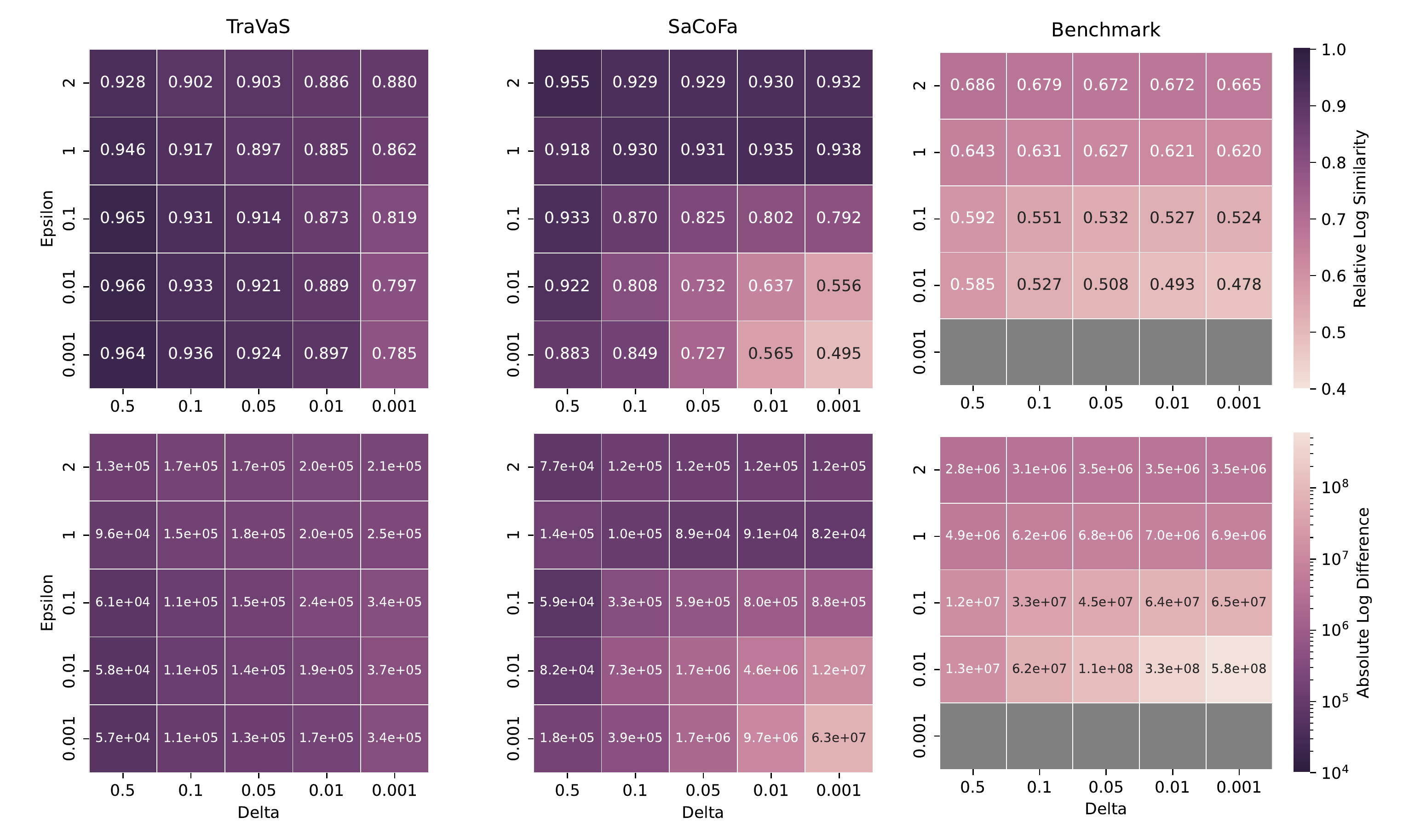}
\caption{The \emph{relative log similarity} and \emph{absolute log difference} results of anonymized BPIC13 logs generated by \textit{TraVaS}, the benchmark, and \textit{SaCoFa}. Each value represents the mean of 10 runs.} \label{fig2}
\vspace{-0.3cm}
\end{figure}

% \vspace{-0.2cm}
\subsection{Data Utility Analysis}

In this subsection, the results of the two aforementioned data utility metrics are presented for both real-life event logs.
We compare the performance of \textit{TraVaS} against our benchmark and \textit{SaCoFa} based on the following privacy parameter values: $\epsilon \in \{2, 1, 0.1, 0.01, 0.001\}$ and $\delta \in \{0.5, 0.1, 0.05, 0.01, 0.001\}$.

Figure~\ref{fig2} shows the average results on BPIC13 in a four-fold heatmap. The grey fields represent a general unfeasibility of the strong privacy setting $\epsilon {=} 0.001$ for our benchmark method. Due to the intense noise perturbation, the corresponding variant generation process increased the number of artificial variant fluctuations to an extent that could not be averaged in a reasonable time.
Apart from this artifact, both \textit{relative log similarity} and \textit{absolute log difference} show superior performance of \textit{TraVaS} for most investigated $(\epsilon,\delta)$ combinations.
In particular, for stronger privacy settings, \textit{TraVaS} provides a significant advantage over \emph{SaCoFa} and benchmark.
Whereas more noise, i.e., lower $(\epsilon,\delta)$ values, generally decreases the output similarity to the original data, \textit{TraVaS} results seem to particularly depend on $\delta$. 
According to Definition~\ref{def:dp_partition}, this observation can be explained by the stronger relation between $k$ and $\delta$ compared to $k$ and $\epsilon$.

%As a result, \emph{TraVaS} may be advantageous in real-world applications since it enables stronger privacy guarantees without significantly degrading data utility by allowing lower $\epsilon$ values at a fixed $\delta$. 

\begin{figure}[t]
\centering
\includegraphics[width=0.95\textwidth]{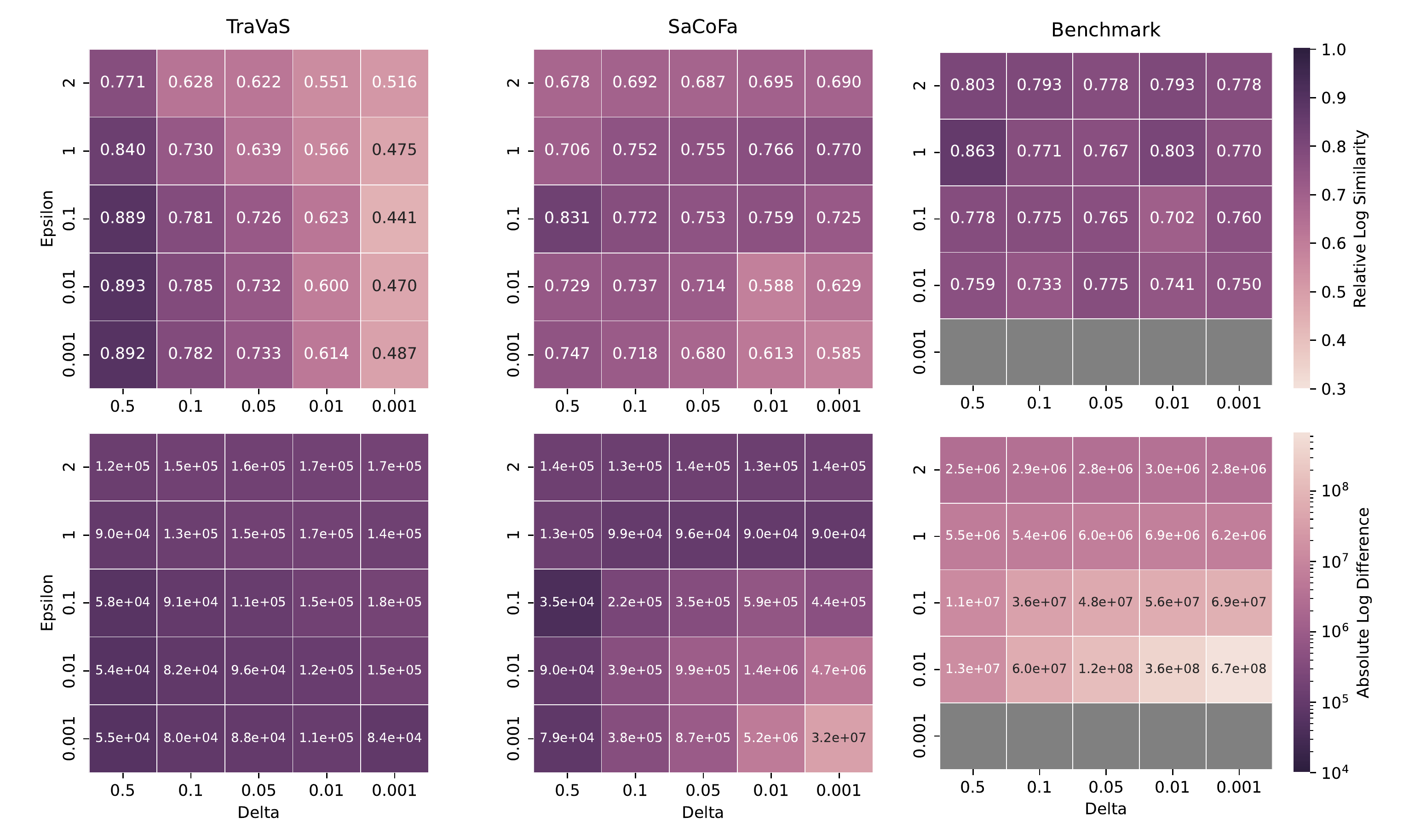}
\caption{The \emph{relative log similarity} and \emph{absolute log difference} results of anonymized Sepsis event logs generated by \textit{TraVaS}, the benchmark, and \textit{SaCoFa}. Each value represents the mean of 10 runs.} \label{fig4}
\vspace{-0.3cm}
\end{figure}

The evaluation of the Sepsis log is presented in Fig.~\ref{fig4}. In contrast to BPIC13, Sepsis contains many variants occurring only once or twice.
While our \emph{absolute log difference} shows a similar expected trend with $(\epsilon,\delta)$ as Fig.~\ref{fig2}, the \emph{relative log similarity} metric indicates almost constant values for the prefix-based techniques and a considerable $\delta$-dependency for \textit{TraVaS}. We explain the resulting patterns by examining the underlying data structure in more detail. 
As mentioned, the frequency threshold $k$ of \textit{TraVaS} strongly correlates with $\delta$. Hence, event logs with prominent infrequent traces, e.g., Sepsis, are significantly truncated for strong $(\epsilon,\delta)$-DP. Since this variant removal leads to a distribution mismatch when being compared to the original log, the \textit{relative log similarity} forms a step-wise pattern as in Fig.~\ref{fig4}.
In contrast, the prefix-based techniques iteratively generate variants that may or may not exist in the original log.
% For event logs with high trace uniqueness, this process adds similar layers of noisy variant distributions that do not significantly change with different privacy guarantees. Hence, we obtain similar relative frequencies and thus data utilities.
In logs with high trace uniqueness, there exist many unique variants that are treated similarly to non-existing variants due to close frequency values, i.e., 0 and 1. Thus, in the anonymized logs, unique variants either appear with larger noisified frequencies or are replaced with fake variants having larger noisified frequencies. 
This process remains the same for different privacy settings but with larger frequencies for stronger privacy guarantees. 
Hence, the \textit{relative log similarity} metric stays almost constant although the noise increases with stronger privacy settings.
However, the \emph{absolute log difference} metric can show differences.
% To assess data quality on small, strongly trace-unique logs, we therefore recommend to employ an \emph{absolute log difference} metric.
\emph{uTraVaS} shows even better performance w.r.t. the data utility metrics.\footnote{\scriptsize \url{https://github.com/wangelik/TraVaS/tree/main/experiments}}

%On the contrary, our state-of-the-art benchmark relies on iteratively generated variant names that may not exist in the ground truth. These variants equipped with their noisified frequencies are then additionally pruned to reduce the computational complexity. For event logs with high trace uniqueness, this process can replace infrequent traces with new artificial variants with larger frequencies.
%Consequently, the \textit{relative log similarity} stays approximately constant although the introduced noise strongly increases with stricter privacy settings (upper right of Fig.~\ref{fig4}).
% Thus, to evaluate the general data distortion induced by $(\epsilon,\delta)$-DP in the \textit{Sepsis} example, it is better to consult an \emph{absolute log difference} metric instead of a distribution comparison (lower parts of Fig.~\ref{fig4}).

% As expected, \emph{SaCoFa} reflects the same trends on both metrics while showing a slightly worse \emph{relative log similarity} performance and a slightly better \emph{absolute log difference} performance than our benchmark.\footnote{\scriptsize \url{https://github.com/wangelik/TraVaS/blob/main/experiments/Sepsis} } 
% Nevertheless, \emph{TraVaS} still outperforms \emph{SaCoFa} w.r.t. the \emph{absolute log difference}, particularly for stronger privacy settings. 

\begin{figure}[bt]
\centering
\includegraphics[width=0.95\textwidth]{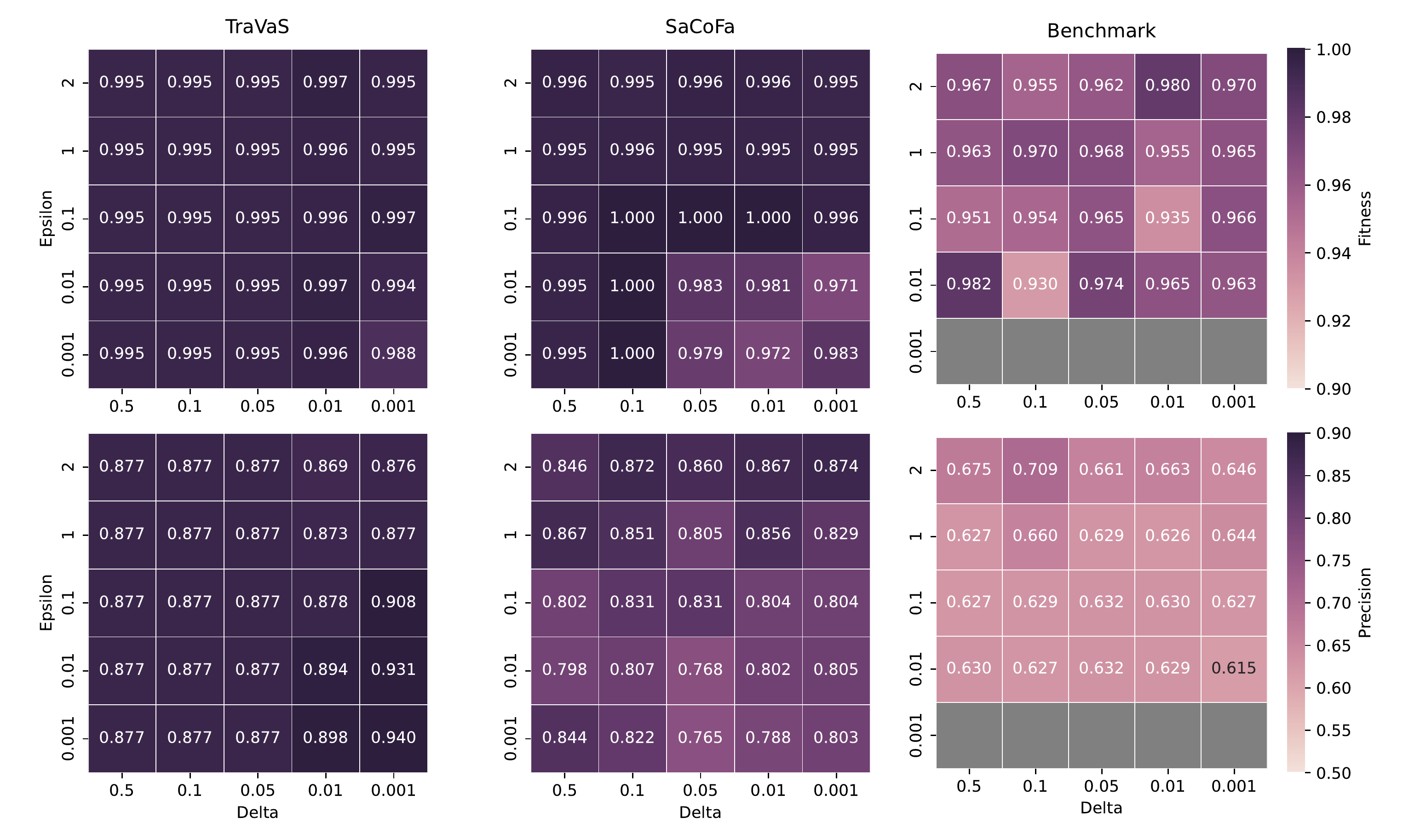}
\caption{The \textit{fitness} and \textit{precision} results of anonymized BPIC13 event logs generated by \textit{TraVaS}, the benchmark, and \textit{SaCoFa}. Each value represents the mean of 10 runs.} \label{fig3}
\vspace{-0.3cm}
\end{figure}

% \vspace{-0.2cm}
\subsection{Process Discovery Analysis}
We conduct a \textit{process discovery} investigation based on \textit{fitness} and \textit{precision} scores. For the sake of comparability, the experimental setup remains unchanged.
Figure~\ref{fig3} shows the results for BPIC13, where the original fitness and precision values are 0.995 and 0.877, respectively.
\textit{TraVaS} provides almost perfect replay behavior w.r.t. \textit{fitness} while the prefix-based alternatives show lower values.
% especially on stronger privacy settings. 
This observation can be explained by the different algorithmic approach of \textit{TraVaS} and some characteristics of BPIC13. \textit{TraVaS} only adopts true behavior that results in a simplified representation of the original process model.
Due to the rather low trace uniqueness and comparably large log-size of BPIC13, this simplification is minor enough to allow an almost perfect fitness. 
In contrast, the fake variants generated by prefix-based approaches negatively affect their fitness scores.
The precision metric evaluates the fraction of behavior in a model discovered from an anonymized log that is not included in the original log. Due to the direct release mechanism of \emph{TraVaS} that only removes infrequent variants, we achieve more precise process models than the alternatives.
Furthermore, the correlation between threshold $k$ and noise intensity enables \emph{TraVaS} to even rise precision for stronger privacy guarantees. Conversely, the fake variants generated by prefix-based approaches can lead to inverse behavior.
%For all $(\epsilon,\delta)$ combinations, we again outperform the benchmark. Here, the reason is rooted in a lower impact of the \textit{TraVaS} thresholding with $k$ on true frequent behavior that are covered by the model obtained from the original log.

\begin{figure}[bt]
\centering
\includegraphics[width=0.95\textwidth]{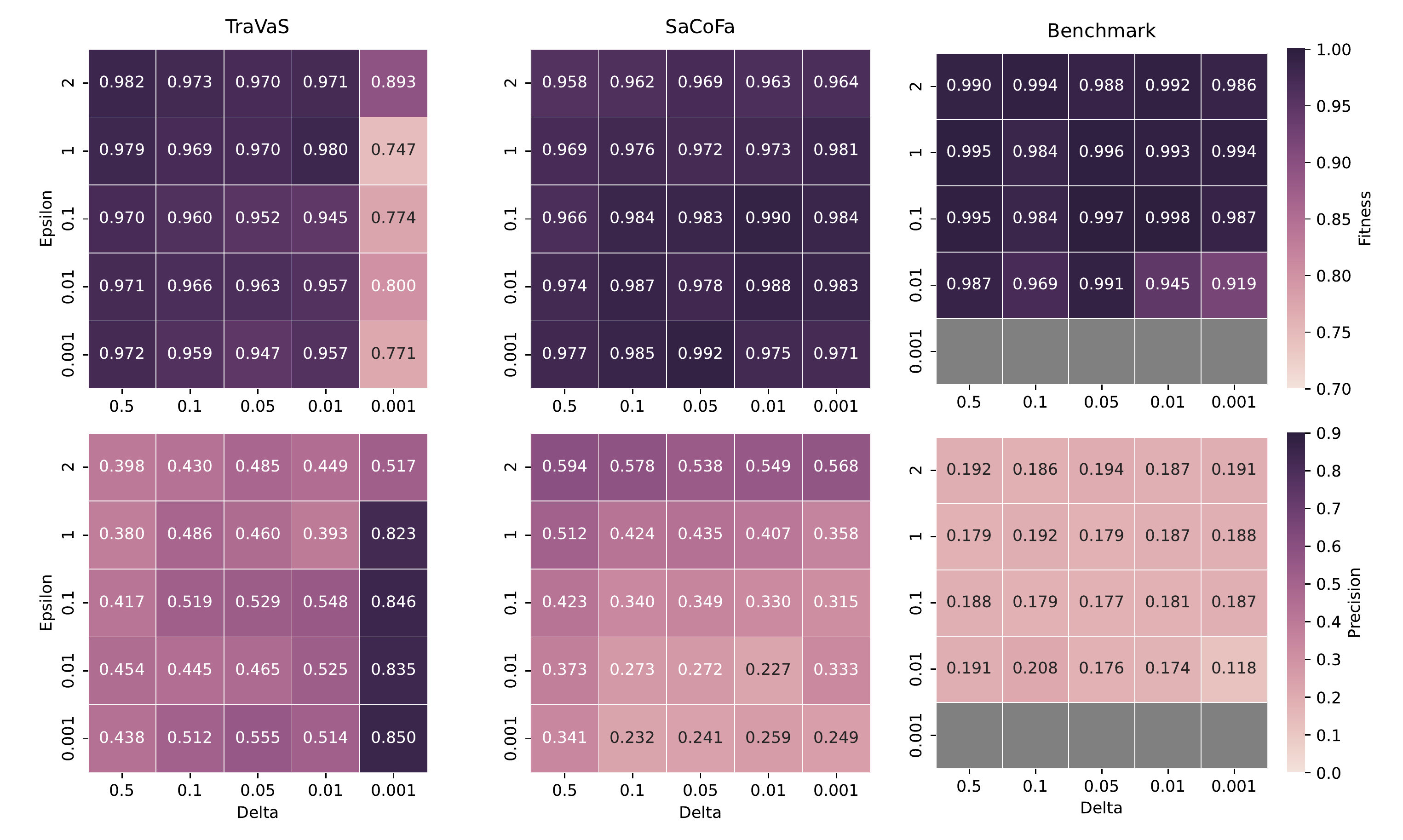}
\caption{The \textit{fitness} and \textit{precision} results of anonymized Sepsis event logs generated by \textit{TraVaS}, the benchmark, and \textit{SaCoFa}. Each value represents the mean of 10 algorithm runs.} \label{fig5}
\vspace{-0.3cm}
\end{figure}

Figure~\ref{fig5} shows the \emph{fitness} and \emph{precision} results for Sepsis, where the original fitness and precision values are 0.952 and 0.489, respectively.
Whereas \textit{TraVaS} dominates the prefix-based approaches w.r.t. \textit{precision} as in Fig.~\ref{fig3}, our fitness score shows a slight under-performance.
Unlike BPIC13, the high trace uniqueness and smaller log-size prohibit the underlying \textit{partition selection} mechanism to achieve negligible threshold for infrequent variant removal.
Thus, the discovered process models from anonymized logs miss parts of the original behavior. 
This shows that carefully tuned prefix-based mechanisms might have an advantage in terms of fitness for small logs with many unique traces.
We particularly note that this limitation of \emph{TraVaS} vanishes as soon as the overall log-size grows. The reason lies in the size-independent threshold $k$ while the pruning parameter of prefix-based approaches intensifies with the data size.
The process discovery analyses for \emph{uTraVaS}, available on GitHub, show even better performance.
% \footnote{\scriptsize \url{https://github.com/wangelik/TraVaS/tree/main/experiments}}

\section{\label{c.conc}Discussion and Conclusion}

%With the rising applicability of process mining and event data analysis, there have emerged various scenarios demanding privacy guarantees during process modeling, data transfer, or even raw storage.
%Conventional state-of-the-art solutions primarily anonymize event data by introducing noise according to specific $(\epsilon,\delta)$-DP mechanisms. These mechanisms perturb the occurrence numbers of trace variants and hide their existence based on suitable trace guessing queries.
%Unfortunately, the underlying guessing routines are computationally expensive, introduce numerous artificial false variants and depend on many hyperparameters such as a maximal variant length or frequency pruning.

In this paper, we demonstrated a novel approach to release anonymized distributions of trace variants based on $(\epsilon,\delta)$-DP mechanisms. The corresponding algorithm (\textit{TraVaS}) overcomes the variant generation problems of prefix-based mechanisms (see Section~\ref{c.intro}) and directly queries all true variants. 
% $(\epsilon,\delta)$-DP of \textit{TraVaS} is ensured by a model-inherent combination of noise addition and frequency thresholding.
Our experiments with two differently structured event logs showed that \textit{TraVaS} outperforms the state-of-the-art approaches in terms of \textit{data utility} metrics and process-discovery-based \textit{result utility} for most of the privacy settings. In particular, for large event logs containing many long trace variants, our implementation has no efficient alternative.
Regarding limitations and future improvements, we generally note that the differentially private partition selection mechanism only works for $\delta {>} 0$, whereby limits of small values can be problematic on large collections of infrequent variants. Thus, all use cases that require strict $\epsilon$-DP still need to apply prefix-based mechanisms. Finding a more efficient solution for $\delta {=} 0$ seems to be a valuable and interesting future research topic.

%
% ---- Bibliography ----
%
% BibTeX users should specify bibliography style 'splncs04'.
% References will then be sorted and formatted in the correct style.
%
\bibliographystyle{splncs04}
\bibliography{References}

\newcommand{\SortNoop}[1]{}
\begin{thebibliography}{10}
\providecommand{\url}[1]{\texttt{#1}}
\providecommand{\urlprefix}{URL }
\providecommand{\doi}[1]{https://doi.org/#1}

\bibitem{GDPR0}
{GDPR}, \url{http://data.europa.eu/eli/reg/2016/679/oj}, {A}ccessed: 2021-05-15

\bibitem{book_wil}
van~der Aalst, W.M.P.: Process Mining - Data Science in Action, Second Edition.
  Springer (2016)

\bibitem{PSO}
Cohen, A., Nissim, K.: Towards formalizing the gdpr's notion of singling out.
  Proc. Natl. Acad. Sci. {USA}  \textbf{117}(15),  8344--8352 (2020)

\bibitem{priv_part}
Desfontaines, D., Voss, J., Gipson, B., Mandayam, C.: Differentially private
  partition selection. Proc. Priv. Enhancing Technol.  \textbf{2022}(1),
  339--352 (2022)

\bibitem{data_bpic}
van Dongen, B.F., Weber, B., Ferreira, D.R., Weerdt, J.D.: {BPI} challenge
  2013. In: Proceedings of the 3rd Business Process Intelligence Challenge
  (2013)

\bibitem{priv_dwork2}
Dwork, C.: Differential privacy: {A} survey of results. In: Agrawal, M., Du,
  D., Duan, Z., Li, A. (eds.) Theory and Applications of Models of Computation,
  5th International Conference. Springer (2008)

\bibitem{priv_graph}
Elkoumy, G., Pankova, A., Dumas, M.: Privacy-preserving directly-follows
  graphs: Balancing risk and utility in process mining. CoRR
  \textbf{abs/2012.01119} (2020)

\bibitem{priv_pripel}
Fahrenkrog{-}Petersen, S.A., van~der Aa, H., Weidlich, M.: {PRIPEL:}
  privacy-preserving event log publishing including contextual information. In:
  Business Process Management - 18th International Conference, {BPM}. Springer
  (2020)

\bibitem{priv_sacofa}
Fahrenkrog{-}Petersen, S.A., Kabierski, M., R{\"{o}}sel, F., van~der Aa, H.,
  Weidlich, M.: Sacofa: Semantics-aware control-flow anonymization for process
  mining. In: 3rd International Conference on Process Mining, {ICPM}. {IEEE}
  (2021)

\bibitem{mine_infreq}
Leemans, S.J.J., Fahland, D., van~der Aalst, W.M.P.: Discovering
  block-structured process models from event logs containing infrequent
  behaviour. Springer (2013)

\bibitem{data_sepsis}
Mannhardt, F.: {Sepsis Cases}  (2016).
  \doi{10.4121/uuid:915d2bfb-7e84-49ad-a286-dc35f063a460}

\bibitem{priv_pinq}
Mannhardt, F., Koschmider, A., Baracaldo, N., Weidlich, M., Michael, J.:
  Privacy-preserving process mining - differential privacy for event logs. Bus.
  Inf. Syst. Eng.  \textbf{61}(5),  595--614 (2019)

\bibitem{priv_emd}
Rafiei, M., van~der Aalst, W.M.P.: Towards quantifying privacy in process
  mining. In: Process Mining Workshops - {ICPM} 2020 International Workshops.
  Lecture Notes in Business Information Processing, Springer (2020)

\bibitem{rafiei_group}
Rafiei, M., van~der Aalst, W.M.P.: Group-based privacy preservation techniques
  for process mining. Data Knowl. Eng.  \textbf{134},  101908 (2021)

\bibitem{cnf}
Tomlin, J.A.: Minimum-cost multicommodity network flows. Oper. Res.  (1966)

\end{thebibliography}

\end{document}